\documentstyle[12pt]{article}
\newcommand{\doublespace}{\renewcommand{\baselinestretch}{1.75}
   \Large\normalsize}

\renewcommand{\ref}[1]{\raisebox{.6ex}{[#1]}}

\newcommand{\be}{\begin{equation}}
\newcommand{\ee}{\end{equation}}

\newcommand{\ol}{\overline }

\begin{document}

\doublespace

\title{ Tunneling with the Lorentz Force and the Friction }

\author{Ping Ao    \\
Department of Theoretical Physics \\
Ume\aa{\ }University, s-901 87, Ume\aa, Sweden  }

\maketitle

\begin{abstract}
We present a semiclassical study of a transport process, the tunneling, 
in the presence of a magnetic field and a dissipative environment.
We have found that the problem can be mapped onto 
an effective one-dimensional one, and the tunneling rate 
is strongly affected by the magnetic field, such as a complete suppression 
by a large parallel magnetic field, an example of the dynamical localization.
In such case a small perpendicular component of the field, 
or the dissipation, can enhance the tunneling rate.
In the small parallel field and finite temperatures 
the tunneling rate is finite.
Explicit expressions will be presented in those cases.
If viewing the tunneling in the presence of a magnetic field as a
dissipative tunneling process, 
by varying the magnetic field and the potential one can obtain the
dissipative spectral function between the subohmic $s =0$ and 
the superohmic $s = \infty$.
In combination with a real dissipative spectral function, the
effect of the magnetic field can map the spectral function from
$s $ to $2-s$, with $s>2$ mapping to $ s = 0$, revealing a dual symmetry
between the friction and the Lorentz force.
Two cases relevant to experiments, the edge state tunneling in a Hall bar 
and the tunneling near the dynamical localization will be discussed in detail.
\end{abstract}


\newpage

\section{Introduction}

Quantum dissipative tunneling has received an extensive study 
recently\cite{caldeira,leggettrmp}, 
because of its important role played in physics, chemistry, 
as well as in biology.
In the study of the quantum transport in semiconductors, 
such as in heterojunctions, 
superlattices, and the quantum Hall systems, this problem
surfaces again, along with a new feature in addition to the  usual 
dissipative effect:
the effect of the magnetic field.\cite{jain,mesoscopic}
This  new feature makes the tunneling behavior much more rich.
The purpose of the present paper is to explore its various 
consequences.\footnote{ to appear in Physica Scripta }

Classically, the equation of motion for a charge particle with mass $m$ 
and charge $q$ is
\be
   m \ddot{\bf R} = - \nabla U({\bf R}) - \eta \dot{\bf R}
                    + \frac{q}{c} \dot{\bf R}\times {\bf B} 
                    + {\bf f}  \; , 
\ee
with the potential $U$, the friction coefficient $\eta$, the magnetic
field ${\bf B}$, and the speed of light $c$.
The fluctuating force is connected to the friction by the 
fluctuation-dissipation theorem.
Eq.(1) is the formiliar Langevin equation, and the friction here
corresponds to the ohmic damping case.
We notice here that both the damping and the magnetic field give
velocity dependent forces which are perpendicular to each other.
This suggests a certain symmetry between the the friction and the Lorentz 
force, which can indeed be borne out as will be shown below.

To study the quantum tunneling, the Hamiltonian of the whole system is needed.
For the Lorentz force, it is straightforward.
The corresponding Hamiltonian is 
\[
   H_B = \frac{1}{2m} \left[ {\bf P} - \frac{e}{c} {\bf A}({\bf R}) 
          \right]^{2} \; ,
\]
with the vector potential ${\bf A}$ determined by 
$ \nabla\times {\bf A} = {\bf B} $.
The formulation of the dissipation within quantum mechanics is not so
straightforward. In the present paper 
the approach and notation of Ref.\cite{caldeira} to this question will
be followed. In that formulation, the dissipation is produced
by a heat bath consisting of many harmonic oscillators, and the corresponding
Hamiltonian is
\[
   H_D = \sum_{j; l=x,y,z} 
       \left[ \frac{1}{2m_{j} } p_{l,j}^{2} + \frac{1}{2} m_{j}
       \omega_{j}^{2} \left(  q_{l,j} - \frac{c_{l,j} }{m_{j}\omega_{j}^{2} }
                        { R_l } \right)^{2} \right] \; .
\]
Here $ (R_x, R_y, R_z) = {\bf R} = (x,y,z)$. 
The effect of the dissipative 
environment is specified by the spectral function
\be
   J_l(\omega) \equiv \pi \sum_{j} \frac{c_{l,j}^{2} }{2m_{j}\omega_{j} } 
               \delta(\omega - \omega_{j} ) \; ,
\ee
with $l= x, y, z$.
The introduction of the notation $l$ stands for the case of a possible
anisotropic dissipative environment.
In the present paper, we shall assume the spectral function to have the 
following form
\be
   J_l(\omega) = \eta_l \omega^{s_l} 
         \exp\left\{ - \frac{\omega}{\omega_{c} } \right\} \; ,
\ee 
with $\omega_{c}$ the cutoff frequency whenever needed.
In accordance with Ref.\cite{caldeira},
$s_l> 1 $ is the superohmic case, $s_l=1$ the ohmic case, and $0\leq s_l<1$ 
the subohmic case. 
In the ohmic damping case, $\eta_l$ is the friction coefficient in Eq.(1).

Now, the total Hamiltonian of our system has the following form:
\be
   H =   \frac{1}{2m} \left[ {\bf P}-\frac{e}{c} {\bf A}({\bf R}) \right]^{2}
       + U({\bf R})  
   + \sum_{j; l=x,y,z} 
       \left[ \frac{1}{2m_{j} } p_{l,j}^{2} + \frac{1}{2} m_{j}
       \omega_{j}^{2} \left(  q_{l,j} - \frac{c_{l,j} }{m_{j}\omega_{j}^{2} }
                        { R_l } \right)^{2} \right] \; .
\ee
We further assume the scalar 
potential $U$ for the charged particle to have the form:
\be
    U = \frac{1}{2} m \omega_{x}^{2} x^{2} + 
        \frac{1}{2} m \omega_{y}^{2}y^{2}  + V(z) \; ,
\ee
with $V$ has a metastable point at $z=0$. For large enough 
$z$, $V(z) = - F z$, and there is a barrier separating this region and $z=0$.
The choice of the potential in the $x-y$ plane is to mimic the impurity
potential.
The external magnetic field is tilted: 
${\bf B} = (B_{\parallel}, 0 , B_{\perp} )$.
Here the magnetic field parallel to the $x-y$ plane is taken along the 
$x$-direction.
In accordance with the calculation of tunneling, 
the vector potential will be taken as
\be
   {\bf A} =  (0, B_{\perp} x - B_{\parallel} z, 0) \;,
\ee
and it can be shown that results are independent of the choice of gauge,
because of the periodic boundary condition in the calculation of the tunneling
rate.
It is clear now we have encountered the problem of tunneling 
in a higher dimensional space: 
the particle moves in 3-d, and the effective dimension for the
dissipative environment is infinite.
The suitable tool to deal with such a
problem is the path integral method, which will be used below.

The organization of the rest paper is as follows.
In the next section using the path integral method 
we reduce the infinite dimensional problem to
an effective one dimensional one, with normal and anomalous damping kernels.
The situation corresponding to the geometry of a Hall bar will be discussed
in section 3. Here the dynamical localization induced by the parallel
magnetic field will be discussed, as well as the symmetries between 
the damping and the magnetic field.
In section 4 the tunneling in the presence of a tilted magnetic field 
will be discussed, with the emphasis on the behaviors near the 
dynamical localization induced by a parallel magnetic field.
We briefly summarize in section 5.

\section{Effective Action }

In this section we use the path integral method to reduce the action
in the infinite dimension to an effective one dimensional one. 

The tunneling is described by the Euclidean 
action
\[
   S = \int_{0}^{\hbar\beta } d\tau \left[ \frac{1}{2} m \dot{\bf R}^{2} + 
      i \frac{q}{c} ( B_{\perp} x - B_{\parallel} z )\dot{y} + 
       V(z) + \frac{1}{2} m \omega_{x}^2 x^{2} 
            + \frac{1}{2} m \omega_{y}^2 y^{2}
        \right.
\]
\be
    +  \sum_{j; l=x,y,z} 
       \left. \left( \frac{1}{2} m_{j} \dot{q}_{l,j}^{2} 
              + \frac{1}{2}m_{j}\omega_{j}^{2} \left( q_{l,j} 
                           - \frac{c_{l,j} }{m_{j} \omega_{j}^{2} } 
                       R_l \right)^{2} \right) \right] \; ,
\ee
where $\beta = 1/k_{B}T$ is the inverse temperature.
The tunneling rate is equal to $\exp\{ - S_{c}/\hbar \} $, where 
the semiclassical action $S_{c}$ is determined by the 
bounce solution of the equation $\delta S =0$, in which the periodic boundary 
condition, $({\bf r}(\hbar\beta), \{q_{j}(\hbar\beta)\}) = ({\bf r}(0), 
\{q_{j}(0)\} )$, is required.

This is so-called the method of the imaginary part of the free 
energy $F = -k_{B}T \ln Z$ calculated from the partition function
\be
   Z = \int d{\bf R}'d\{{\bf q}_j'\} \; 
              D({\bf R}', \{{\bf q}_j'\}; {\bf R}', \{{\bf q}_j'\} ) \; , 
\ee
with the density matrix 
\be
   D({\bf R}'(0), \{{\bf q}_j'(0)\} ; 
           {\bf R}''(\hbar\beta), \{{\bf q}_j''(\hbar\beta)\}) 
   = \int {\cal D}{\bf R} {\cal D}\{{\bf q}_j\} \; 
     \exp\left\{- \frac{1}{\hbar}  S[{\bf R},\{{\bf q}_j\}]   \right\} \; .
\ee
This method is identical to the WKB method at zero temperature, and allows us 
to have a unified treatment of the escape rate for finite 
temperatures.\cite{hanggi}

We are interested in the particle tunneling out of the metastable state $z=0$ 
in the $z$-direction. 
After the tunneling other degrees of freedoms, $x$ and $y$, 
as well as $\{ q_{l,j} \}$, can take any allowed value. 
Therefore the summation over final states, 
integrations over the $x$ and $y$, as well as $\{ q_{l,j} \}$
 coordinates, will be taken in the 
calculation of the partition function. Those are Gaussian integrals.
To integrate over the $x, y$ and $\{q_{l,j}\}$ coordinates, 
we perform a Fourier 
transformation on the time interval $[0, \; \hbar\beta]$:
\be
   (x(\tau), y(\tau), z(\tau) ) = 
             \frac{1}{\hbar\beta} \sum_{n=-\infty}^{\infty}
             (x_{n}, y_{n}, z_{n} ) e^{i \nu_{n} \tau } \; ,
\ee
and 
\be
   q_{l,j}(\tau) = \frac{1}{\hbar\beta} \sum_{n=-\infty}^{\infty}
             q_{l,j, n} e^{i \nu_{n} \tau } \; .
\ee
Here $\nu_{n} = 2\pi n /\hbar\beta \; $.
The action $S$ of Eq.(7) can be then rewritten as
\[
   S[\{ {\bf R}_{n} \}, \{ q_{l,j,n} \} ] =  V(z)_{n}  
      + \frac{1}{\hbar\beta} \sum_{n=-\infty}^{\infty} 
      \left[ \frac{1}{2} m \nu_{n}^{2} z_{n} z_{-n} 
      + \frac{1}{2} ( m \nu_{n}^{2} + m \omega_{x}^{2}  ) x_{n} x_{-n} \right]
\]
\[
   + \frac{1}{\hbar\beta} \sum_{n=-\infty}^{\infty} 
   \left[ \frac{1}{2} ( m \nu_{n}^{2} + m \omega_{y}^{2} ) y_{n} y_{-n} 
     + \frac{e}{c}( B_{\perp} x_{n} - B_{\parallel} z_{n}) 
      \nu_{-n} y_{-n} \right] 
\]
\be
   + \frac{1}{\hbar\beta} \sum_{n=-\infty}^{\infty} \sum_{l,j}
   \left[ \frac{1}{2} m_j \nu_{n}^{2} q_{l,j,n} q_{l,j,-n}
   +   \frac{1}{2} m_j \omega_{j}^{2}
    \left(q_{l,j,n} - \frac{c_{l,j}}{m_j\omega_j^2 } R_{l,n} \right)
    \left(q_{l,j,-n}-\frac{c_{l,j}}{m_j\omega_j^2 } R_{l,-n}\right)\right] \; ,
\ee
with 
\be 
   V(z)_{n} \equiv \hbar\beta\int^{\hbar\beta}_0 d\tau V(z(\tau)) \; .
\ee
Integration over $\{ q_{l,j,n}\}$, we have 
\[
   S_{eff}[\{ {\bf R}_{n} \} ] = V(z)_{n} +
      \frac{1}{\hbar\beta} \sum_{n=-\infty}^{\infty} 
      \left[ \frac{1}{2} (  m \nu_{n}^{2} + \xi_{z,n} ) z_{n} z_{-n} 
      + \frac{1}{2} ( m \nu_{n}^{2} + m \omega_{x}^{2} + \xi_{x,n} )
          x_{n} x_{-n} \right]
\]
\[
   + \frac{1}{\hbar\beta} \sum_{n=-\infty}^{\infty} 
   \left[ \frac{1}{2} ( m \nu_{n}^{2} + m \omega_{y}^{2} + \xi_{y,n} ) 
      y_{n} y_{-n} 
     + \frac{e}{c}( B_{\perp} x_{n} - B_{\parallel} z_{n}) 
      \nu_{-n} y_{-n} \right] \; ,
\]
with 
\be 
   \xi_{l,n} = \frac{1}{\pi} \int_0^{\infty} d\omega  J_l(\omega)
               \frac{2 \nu_n^2 }{ \nu_n^2 + \omega^2 } \; .
\ee
Here $l= x,y,z$.
Similarly, the integration over $\{x_n, y_n \}$ can be done.
The resulting effective action is
\be
      S_{eff}[ z(\tau) ] = \int_{0}^{\hbar\beta } d\tau 
    \left[ \frac{1}{2} m \dot{z}^{2} + V(z) \right]
    + \frac{1}{2} \int_{0}^{\hbar\beta } d\tau \int_{0}^{\hbar\beta } d\tau'
      [ k(\tau - \tau' ) 
      + g(\tau - \tau' ) ] [ z(\tau) - z(\tau') ]^{2} \; ,
\ee
with the normal damping kernel $k$ as
\be 
   k(\tau) = \frac{1}{\pi} \int_{0}^{\infty} d\omega \;  J(\omega) 
             \frac{ \cosh[\omega (\frac{\hbar\beta}{2} - |\tau| ) ] }
                  { \sinh[\frac{\omega\hbar\beta}{2} ] } \; ,
\ee
and the damping kernel due to the magnetic field mixed up $x-y$ direction 
motion contributions, 
which we shall call the anomalous damping kernel, as
\[
   g(\tau) = - 
         \frac{1}{\hbar\beta} 
     \sum_{n=-\infty}^{\infty} \frac{1}{2} 
     \frac{ (\frac{e}{c} B_{\parallel} )^{2} \nu_{n}^{2} }
          { m\nu_{n}^{2} + m\omega_{y}^{2} + \xi_{y,n} } \;
             e^{ i\nu_{n} \tau } 
\]
\be
   +  \frac{1}{\hbar\beta} \sum_{n=-\infty}^{\infty} \frac{1}{2} 
         \frac{\left[ B_{\perp} B_{\parallel} ( \frac{e}{c} )^{2} 
                 \nu_{n} \nu_{-n}  \right]^{2}         }
             {  [ m\nu_{n}^{2} + m\omega_{y}^{2} + \xi_{y,n} ] 
          \left[ \left(\frac{e}{c} B_{\perp} \right)^{2} \nu_{n}^{2} 
            + ( m\nu_{n}^{2} + m\omega_{x}^{2}+ \xi_{x,n}  )
              ( m\nu_{n}^{2} + m\omega_{y}^{2}+ \xi_{y,n}  )\right] } \; 
             e^{ i \nu_{n} \tau }        \; .
\ee
In obtaining the normal damping kernel $k$ we have used 
$ \sum_n \frac{\omega^2 }{\nu_n^2 + \omega^2 } e^{i\nu_n\tau}
 = \frac{\hbar\beta}{2} \frac{\cosh[\omega(\hbar\beta/2 - |\tau|)] }
                             {\sinh[\omega\hbar\beta/2] }$,
and dropped a term containing the periodic delta function 
$\sum_{n=-\infty}^{\infty} e^{ i \nu_{n} \tau }$.
In the large $\tau$ limit, from Eqs.(2) and (16) the normal damping kernel 
takes the form
\be
   k(\tau) = \frac{1}{\pi} \; \eta \; \frac{1}{\tau^{s+1} } \; ,
\ee
which demonstrates the one-to-one correspondence between the 
low frequency part 
of the spectral function and the long time behavior of the damping kernel.

Now we have obtained an effective one-dimensional problem. 
The physics due to the influence of the damping and the magnetic 
field is buried deeply in the normal and anomalous damping kernels.
In the following we study two special geometric arrangements to reveal
the implications.

\section{Edge State Tunneling }  
 
In this section we study the tunneling of the charged particle 
confining to two dimensions, that is, 
there is no perpendicular magnetic field perpendicular to the $x-y$ plane, 
and the $x$-direction motion of the particle is irrelavent.

The effective action is given by Eq.(15), with the normal damping kernel of 
Eq.(16). 
In the absence of the perpendicular magnetic field, $ B_{\perp} = 0$, 
the second term in Eq.(17) is zero.
The anomalous damping kernel $g$ is 
\[
     g(\tau) = - 
         \frac{1}{\hbar\beta} 
     \sum_{n=-\infty}^{\infty} \frac{1}{2} 
     \frac{ (\frac{e}{c} B_{\parallel} )^{2} \nu_{n}^{2} }
          { m\nu_{n}^{2} + m\omega_{y}^{2} + \xi_{y,n} } \;
             e^{ i\nu_{n} \tau } \; .
\]
This kernel can be split into two terms,
\[
     g(\tau) = - 
         \frac{1}{\hbar\beta} \left(\frac{e}{c} B_{\parallel} \right)^{2}
     \left[\sum_{n=-\infty}^{\infty} \frac{1}{2} e^{ i\nu_{n} \tau } -
     \sum_{n=-\infty}^{\infty} \frac{1}{2} 
     \frac{ m\omega_{y}^{2} + \xi_{y,n} }
          { m\nu_{n}^{2} + m\omega_{y}^{2} + \xi_{y,n} } \;
             e^{ i\nu_{n} \tau }\right] \; .
\]
The first term is a delta function, which gives no contribution to the 
semiclassical action under the periodic boundary condition. 
We only keep the second term:
\be
    g(\tau) =  
         \frac{1}{\hbar\beta} \left(\frac{e}{c} B_{\parallel} \right)^{2}
     \sum_{n=-\infty}^{\infty} \frac{1}{2} 
     \frac{ m\omega_{y}^{2} + \xi_{y,n} }
          { m\nu_{n}^{2} + m\omega_{y}^{2} + \xi_{y,n} } \;
             e^{ i\nu_{n} \tau }  \; .
\ee 
Thus the anomalous damping kernel takes a particular simple form.
This allows us for a more detailed calculation of the semiclassical
action.

\subsection{ Parallel Magnetic Field Only }
 
This is the case of the tunneling between two parallel equal potential states
in the presence of a magnetic field, 
a situation identical to the edge state 
tunneling in the Hall bar.
The magnetic field, $B_{\parallel}$, is parallel to the $x-y$ plane but
perpendicular to the plane of the Hall bar.
Similar cases have been studied in vortex tunneling in superconductors and
and superfluids.\cite{aovo,blatter}

From Eq.(19) in the limit of $\omega_{y} = 0$, only the $n=0$ mode
has a finite contribution.
Then the effective action is
\be
   S_{eff} = \int_{0}^{\hbar\beta } d\tau \left[ \frac{1}{2} 
      m \dot{z}^{2} + V(z)  \right] 
    + \frac{1}{4m} \left(\frac{q}{c} B_{\parallel} \right)^{2} 
      \frac{1}{\hbar\beta} 
      \int_{0}^{\hbar\beta} d\tau \int_{0}^{\hbar\beta} d\tau' 
      [ z(\tau) - z(\tau') ]^{2} \; .
\ee
We note that in the language for the spectral function of the dissipative 
environment 
this corresponds to the     
subohmic bath case with $s=0$({\it c.f.} Eqs.(2) and (18)), and 
Eq.(20) is explicitly gauge invariant under the change $z \rightarrow
z + \; constant$, because of the periodic boundary condition of $y$ imposed in 
the tunneling calculation.

For a small enough magnetic field  $B_{\parallel}$,  
there is a tunneling solution for the semiclassical equation 
$\delta S_{eff}[q] = 0$. Particularly, 
for a very small $B_{\parallel}$ and low temperatures 
the semiclassical action may be evaluated perturbatively:
\be
   S_{c} = \int_{-\infty}^{\infty } d\tau \left[ \frac{1}{2} 
      m \dot{z}^{2}_{c}(\tau) + V( z_{c}(\tau) )  \right] 
    + \frac{1}{2m} \left( \frac{q}{c} B_{\parallel} \right)^{2} 
      \left[ \int_{-\infty}^{\infty} d\tau  z_{c}^{2}(\tau) 
    - \frac{1}{\hbar\beta} \left( \int_{-\infty}^{\infty} 
                                 d\tau  z_{c}(\tau) \right)^{2} \right] \; ,
\ee   
where $z_{c}(\tau)$ is the bounce solution at zero temperature without 
the magnetic field.
Eq.(21) shows a remarkable $B_{\parallel}^{2}$ dependence and 
linear temperature dependence.
This feature has been verified experimentally.\cite{aomd}

For large enough magnetic fields, 
the tunneling rate vanishes at zero temperature. 
One way to understand this result is that  the magnetic field effectively 
renormalizes the original potential such that the state near $z=0$ becomes 
stable. This can be noted from Eq.(20) where as $T \rightarrow 0$ 
the cross term in the last term vanishes linearly in $T$, 
and results in an additional
term in the potential. The effective action has the following form in this
case:
\be 
  S_{eff} = \int_{-\infty}^{\infty } d\tau \left[ \frac{1}{2} 
      m \dot{z}^{2} + V(z)  
    + \frac{1}{2m} \left(\frac{q}{c} B_{\parallel} \right)^{2} 
        z^2(\tau) \right] \; .
\ee 
A straightforward calculation leads to the criterion for the 
localization as
\be
    \frac{1}{m} \left( \frac{q}{c} B_{\parallel}^{2}\right)^{2} > 
     \left|\frac{d^{2} V(z) }{dz^{2} } \right|_{barrier\; top} \; .
\ee
Thus we have obtained that the $s=0$ dissipative environment is marginal 
for localization in tunneling decay, compared to the $s=1$ case for the 
tunneling splitting\cite{leggettrmp}.
This is the dynamical localization caused by the parallel magnetic field
\cite{aomd}.
It is worthwhile to point out that according to Eq.(23) 
although a large magnetic inhibits tunneling, 
a large particle mass instead favors the tunneling, contrast to
the usual tunneling process where a large particle mass disfavors
the tunneling.
 
An alternative understanding of this dynamical 
localization is to  imagine that the particle has a finite cyclotron 
orbit inside the tunneling barrier. Increasing the magnetic corresponds to 
decreasing the orbit size. 
Eventually the orbit is smaller than the barrier region, and the particle
is trapped inside the barrier. Tunneling out of the metastable state
becomes impossible.
However, a finite temperature will destroy this localization.
We will return to this point in section 4.

\subsection{ Finite Dissipation}

Now we set the impurity potential $\omega^{2}_{y} = 0$ in Eq.(19). 
Both the normal damping kernel and the anomalous damping kernel are finite.
Since the quantum tunneling corresponds 
to the limit of large imaginary time, $\hbar\beta \rightarrow \infty$,
we look for the large time limit behavior of the anomalous damping kernel $g$.
In the this limit, we may replace 
the summation $1/\hbar\beta \sum_{n}$ by the integration $ 1/2\pi \int d\nu $.
Then for the environment $0 < s < 2$, 
we find the anomalous damping kernel $g$ in the large $\tau$ limit as
\be
   g(\tau) =  a \; (2-s_y) \; \frac{1}{2\pi m} 
                \left( \frac{q}{c} B_{\parallel}\right)^{2} \; 
          \frac{ 1 }{ \ol{\eta} } \; \frac{1}{\tau^{2-s_y + 1} } \; ,
\ee
with $a$ a numerical constant of order of unity.
Here 
\be
   \ol{\eta} = \frac{2}{\pi} \frac{\eta_y}{m} \int_{0}^{\infty} dz 
                     \frac{z^{s_y -1} }{z^{2} + 1 } \; ,
\ee
and $J_y(\omega) = \eta_y \omega^{s_y} $ has been used.
The effective dissipative environment corresponding to the anomalous 
damping kernel is $s_{eff} = 2-s$({\it c.f.} Eqs.(2) and (18)), 
which leaves the ohmic damping unchanged, transforms the subohmic damping 
into superohmic damping and {\it vice versa}.

For the case $s> 2$, using Eq.(3) for the spectral function $J$, 
we find that the effective dissipative environment corresponding 
to the anomalous damping kernel is
$s_{eff} = 0$, which smoothly connects the result for $0 < s < 2$.

An important example is the ohmic damping case.
Carrying out a detailed but straightforward calculation, 
we find the effect action at zero temperature as
\be
   S_{eff} = \int_{-\infty}^{\infty} d\tau 
    \left[ \frac{1}{2} m \dot{y}^{2} + V_{1}(y) \right]
        + \frac{1}{2\pi} \; \eta_{eff} 
         \int^{\infty}_{-\infty} d\tau \int^{\infty}_{-\infty}d\tau' 
         \frac{1}{|\tau-\tau' |^{2} } [ y(\tau) - y(\tau) ]^{2} \; ,
\ee
with the effective damping strength $\eta_{eff}$ as
\be
   \eta_{eff} = \eta_z + \left( \frac{q B_{\parallel} }{c} \right)^{2}
                \frac{1}{\eta_y} \; . 
\ee

The effect of a dissipative environment can now be summarized as follows.
It has been demonstrated in Ref.\cite{leggettrmp} that
the subohmic dissipation has strong effects 
on tunneling, while the superohmic dissipation has weak effects.
From the above analysis 
we have that if the normal damping kernel $k(\tau)$ of Eq.(16) is subohmic, 
the anomalous damping kernel $g(\tau)$ of Eq.(19) is superohmic, 
and {\it vice versa}.
Then according to Ref.\cite{leggettrmp} the effect of the magnetic field, 
represented by the anomalous damping kernel, is weak/strong 
on particle tunneling for the normal subohmic/superohmic damping.
In particular, for normal superohmic damping with $s > 2$, 
the particle tunnels as if there were no effect of dissipation. 
For the normal ohmic damping with $s=1$,
we need to compare the relative strength of the magnetic field and the 
dissipation according to Eq.(27).
Classically, the Lorentz force tends to keep a vortex moving along 
an equal potential contour, but the friction instead 
along the gradient of the potential. 
In general, we can conclude that the dissipation 
tends to suppress the effect of the Lorentz force on the tunneling. 

Eqs.(24,25,27) suggest a dual symmetry 
between the damping and the magnetic field.
Mathematically, this property arises from treating the $y$-direction motion
as an environment for the $z$-direction motion, where $y$ and $z$ can be
regarded as a pair of canonical variables in the presence of the
magnetic field, as suggested by the guiding center dynamics, which will become
more clear in the next subsection.

\subsection{ Impurity Potential }

The normal damping kernel vanishes in this situation. The anomalous damping 
kernel can be expressed by hyperbolic functions. 
We find that the effective action in this case  is 
\[
   S_{eff} = \int^{\hbar\beta}_{0} d\tau 
             \left[ \frac{1}{2} m \dot{z}^{2}+V(z) \right]
   + \frac{1}{4m} \left( \frac{q}{c} B_{\parallel} \right)^{2}  
     \int^{\hbar\beta}_{0} d\tau  \int^{\hbar\beta}_{0} d\tau'  \times
\]
\be
    \frac{\omega_{y}}{2}
    \frac{\cosh[\omega_{y} ( \frac{\hbar\beta}{2} - |\tau-\tau'|)] }
         {\sinh[\frac{\omega_{y}\hbar\beta}{2} ] }
    [ z(\tau) - z(\tau') ]^{2} \; .
\ee
It is the superohmic case with $s_{eff} = \infty$, 
because the effective spectral function has the form 
$\delta(\omega -\omega_{x} )$, which has no low frequency mode. 
Therefore the tunneling rate is nonzero for any magnitude of the magnetic
field. 
This result shows that an impurity has a very strong influence on 
the tunneling in the presence of the Lorentz force,
because the introducing of the impurity potential bends the straight line 
trajectory of a particle and makes the transition 
to other trajectories possible.
We note that by letting $\omega_{x} =0$ we recover Eq.(20), where 
the tunneling rate vanishes for a sufficiently magnetic field.

We can evaluate the semiclassical action perturbatively,
if the magnetic field is weak, or, the impurity potential is strong as done 
in Eq.(21).
In the strong magnetic field limit semiclassical action may be evaluated by a 
variational method similar to Ref.\cite{caldeira} in the case of ohmic damping.
However, we have a more powerful method to perform the calculation, 
because a strong magnetic field freezes the kinetic energy of the particle.
In this case $x$ and $y$ coordinates now form a pair of 
canonically conjugate variables, the guiding center dynamics, 
and the Euclidean action is\cite{jain}
\be
   S_{eff} = \int^{\hbar\beta}_{0} d \tau 
             \left[ - i \frac{q}{c} B_{\parallel } \dot{y} z + 
       V(z) + \frac{1}{2} m \omega_{y}^2 y^{2} \right] \; .
\ee
Following the calculation outlined in Ref.\cite{jain},
we find the semiclassical action in the form
\be
   S_{c} = 2 \frac{q}{c} B_{\parallel }  \;  \; \int^{z_{t}}_{0 } dz 
           \sqrt{  \frac{ 2 V(z) }{m \omega_{y}^2 }  } \; ,
\ee
with $z_{t}$ the turning point determined by the equation $V(y) =V(0)$. 
The result shows that as $ m \omega_{y}^2$ increases, the 
semiclassical action decreases.
Therefore we conclude that the impurity potential in the $y$-direction 
helps the tunneling in $z$-direction.

\subsection{Finite Impurity Potential and Dissipation } 

In general we need to go back to Eqs.(15-17) to study the tunneling under the 
influence of an impurity and dissipation.
However, based on the insight gained by the above analysis
we can draw a general conclusion:
Since both the dissipation and impurity tend to suppress the effect 
of the magnetic field, their total effect will do the same.
Indeed the anomalous damping kernel is always superohmic.
In particular, in the presence of a strong impurity potential 
and ohmic damping, the 
superohmic-like anomalous damping kernel $g$ 
may be ignored compared to the ohmic normal damping kernel $k$.
The effect of $g$ is to renormalize bare parameters such as the mass and the
potential.
Then from Eq.(15) we may have the effective action as
\be
    S_{eff} = \int_{0}^{\hbar\beta } d\tau 
    \left[ \frac{1}{2} m \dot{z}^{2} + V(z) \right] + 
     \frac{\eta_z}{2\pi} \int_{0}^{\hbar\beta } 
      d\tau \int_{0}^{\hbar\beta } d\tau'
      \frac{1}{ |\tau - \tau' |^{2} }  [ z(\tau) - z(\tau') ]^{2}  \; ,
\ee
which looks as if there were no effect of the magnetic field.
This result is in accordance with 
the calculation of  the change of the tunneling behavior
in the presence of random impurities.\cite{shklovskii}

\section{Near the Dynamical Localization}

In the previous section we have found semiclassically 
that for a strong enough parallel 
magnetic field the tunneling rate is zero at zero temperature.
This is a dynamical localization induced by the magnetic field.
Questions remain as regarding to the nature of the transition, and 
the behaviors near this localization. 
In this section we calculate the tunneling rate near the dynamical localization
in the absence of the dissipative environment, but allowing a finite
perpendicular magnetic field and a finite temperature.
We will further assume the impurity potential is zero to make the analytical
analysis transparent. We note that the impurity potential behaves
as a finite perpendicular magnetic field.

Without the impurity potential and the dissipation, the effective action is
\[
    S_{eff}[ z(\tau) ] = \int_{0}^{\hbar\beta } d\tau 
    \left[ \frac{1}{2} m \dot{z}^{2} + V(z) \right] 
   + \frac{1}{2} \int_{0}^{\hbar\beta } d\tau \int_{0}^{\hbar\beta } d\tau'
       \frac{m}{2} \left( \frac{e B_{\parallel} }{mc} \right)^{2} \times
\]
\be
        \frac{ \omega_{\perp} }{2} 
        \frac{ \cosh[ \omega_{\perp} (\frac{\hbar\beta}{2} -|\tau-\tau'| ) ] }
             { \sinh[ \frac{ \hbar \beta \omega_{\perp} }{2} ] }
         [ z(\tau) - z(\tau') ]^{2}    \; ,
\ee
with 
$\omega_{\perp} = B_{\perp} e /mc \; $.
The form  of this action is identical to the one given 
by Eq.(28).
Therefore, the introducing of the perpendicular magnetic field 
immediately renders the tunneling rate finite.
To see the nature of the dynamical localization induce by the magnetic field,
we let $\omega_{\perp} = 0$, and goes back to the renormalized potential at
zero temperature( {\it c.f.} Eq.(29)):
\be
  V_{re}(z) =  V(z)  
    + \frac{1}{2m} \left(\frac{q}{c} B_{\parallel} \right)^{2} 
        z^{2}(\tau) \; .
\ee
The semiclassical action is always finite as long as the metastable
exists. This implies that the dynamical localization is a `first order'
phase transition as varying the magnetic field:
below the critical field $B_c$ the semiclassical action is finite, and
above $B_c$ it is infinite( or a jump to another value if the renormalized 
potential has the stable point not at $z=0$).
Hence the first set of result is
\be
   S_c = \left\{
          \begin{array}{lr}
          S_{c0} &     
            {\ } {\ } {\ } {\ } T=0, B_{\perp} = 0, B_{\parallel} \leq B_c  \\
          \infty &    
            {\ } {\ } {\ } {\ } T=0, B_{\perp} = 0, B_{\parallel} > B_c
          \end{array} \right.  \; .     
\ee
Here $S_{c0}$ is the semiclassical action right at the transition.

For a finite but small temperature and zero perpendicular magnetic 
field, the effective action is 
\[  
  S_{eff}[z(\tau)] = \int_{0}^{\hbar\beta } d\tau \left[ \frac{1}{2} 
      m \dot{z}^{2} + V(z)  \right] 
    + \frac{m}{4} \left( \frac{e B_{\parallel} } {mc } \right)^{2} 
      \frac{1}{\hbar\beta} 
      \int_{0}^{\hbar\beta} d\tau \int_{0}^{\hbar\beta} d\tau' 
      [ z(\tau) - z(\tau') ]^{2} \; .
\]
If $B_{\parallel} \leq B_c$, the temperature can be treated as a perturbation.
The calculation is identical to the one leading to Eq.(21).
If $B_{\parallel} > B_c$, the temperature dependence is singular. 
The semiclassical action may be estimated in the following way.
In this case the renormalized potential has two local minima,
that is, $V_{re}'(z) =0 $ has two solutions (We only consider such a
situation here. For a more complicated potential $V$, more than two
minima is possible.), one at $z=0$ and one at $z_2$ 
determined by the magnetic field.
The second minimum is metastable. The tunneling process can now be viewed 
as two steps. First, by the thermal activation the particle raises its
energy to that of the second minimum. Then it tunnels into the second minimum.
Therefore the semiclassical action has the form of 
$S_{c0} + (V_{re}(z_2)- V_{re}(0))/k_B T$. 
Here we note 
$(V_{re}(z_2)- V_{re}(0))|_{B_{\parallel} = B_c } = 0$.
Near the dynamical transition point, the semiclassical action can
be summarized as
\be
   S_c = \left\{
          \begin{array}{lr}
          S_{c0} &       B_{\perp} = 0, B_{\parallel} \leq B_c  \\
          \frac{ \hbar ( V_{re}(z_2)- V_{re}(0)) }{k_B T } + S_{c0} 
          & B_{\perp} = 0, B_{\parallel} > B_c
          \end{array} \right.  \; .     
\ee
There is an Ahrenius-type temperature dependence, with a small
effective activation energy $V_{re}(z_2) - V_{re}(0)$.

In the case of finite perpendicular magnetic field, $B_{\perp} \neq 0$,
the general expression should follow from Eq.(32).
Near the dynamical localization point, again simple expressions for the
semiclassical action can be obtained.
If the temperature is high such as $k_B T > \hbar \omega_{\perp}/2$, 
the effect of $B_{\perp}$ can be ignored. In this case
the semiclassical action takes the same form as for $B_{\perp} = 0$, and is
\be
   S_c = \left\{
          \begin{array}{lr}
          S_{c0} & 2  k_B T > \hbar \omega_{\perp}, B_{\parallel} \leq B_c  \\
          \frac{ \hbar (V_{re}(z_2) - V_{re}(0)) }{ k_B T } + S_{c0} 
          & 2 k_B T > \hbar \omega_{\perp}, B_{\parallel} > B_c
          \end{array} \right.  \; .     
\ee
In the opposite limit, $k_B T < \hbar \omega_{\perp}/2$, the effective action
may take the following form:
\[
    S_{eff}[ z(\tau) ] = \int_{0}^{\hbar\beta } d\tau 
    \left[ \frac{1}{2} m \dot{z}^{2} + V(z) \right] 
   +   \frac{m}{4} \left( \frac{e B_{\parallel} }{mc} \right)^{2}
      \int_{0}^{\hbar\beta } d\tau \int_{0}^{\hbar\beta } d\tau'
        \frac{ \omega_{\perp} }{2} 
        e^{ - \omega_{\perp}|\tau-\tau'| }
         [ z(\tau) - z(\tau') ]^{2}    \; .
\]
If $B_{\parallel} \leq B_c $, the semiclassical action can be evaluated 
perturbatively.
For  $B_{\parallel} > B_c $, we make following estimation.
Since $\hbar\beta > 2/ \omega_{\perp}$, $\omega_{\perp}$ as a cutoff 
may be introduced. The effective action becomes:
\be
    S_{eff}[ z(\tau) ] = \int_{0}^{\frac{2}{ \omega_{\perp} } } d\tau 
    \left[ \frac{1}{2} m \dot{z}^{2} + V(z) \right] 
   +  \frac{m}{4} \left( \frac{e B_{\parallel} }{mc} \right)^{2}
       \int_{0}^{\frac{2}{ \omega_{\perp} } } d\tau 
       \int_{0}^{\frac{2}{ \omega_{\perp} } } d\tau' 
     \frac{ \omega_{\perp} }{2} 
         [ z(\tau) - z(\tau') ]^{2}    \; .
\ee
This suggests that $\omega_{\perp}$ plays the same role as the temperature, 
and the semiclassical action will have the form 
$S_{c0} + (V_{re}(z_2)- V_{re}(0))/\hbar \omega_{\perp}$. 
We summarize the results in this low temperature limit, 
\be
   S_c = \left\{
          \begin{array}{lr}
          S_{c0} &         2 k_B T < \hbar \omega_{\perp} , 
                           B_{\parallel} \leq B_c  \\
          \frac{2 ( V_{re}(z_2) - V_{re}(0)) }{ \omega_{\perp} } + S_{c0} & 
            2 k_B T < \hbar \omega_{\perp}, B_{\parallel} > B_c
          \end{array} \right.  \; .     
\ee
In the small $\omega_{\perp}$ limit $S_c$ goes to $\infty$ when 
 $B_{\parallel} > B_c$.

We must point out that the calculations in this section are of 
 mean field type. No fluctuations around the mean field solutions
are considered, which may be significant in some situations.

\section{ Summary }

We have performed a systematic study of the influence 
of the dissipation and the magnetic field on tunneling in the 
presence of an impurity potential.
Along the tunneling coordinate the effective action is one dimensional,
but with normal and anomalous damping kernels which are no-local in time.
Two cases, the Hall bar geometry and the behaviors near the 
dynamical localization,
have been investigated in detail in order to reveal
the physics implied in the effective action.
A dual symmetry between the damping and the magnetic field has been found,
corresponding the mapping between different heat baths.
The very pronounced result is the dynamical localization caused
by the parallel magnetic field:
For a strong enough parallel magnetic field the tunneling can be
completely suppressed. 
For the weak parallel magnetic field magnetic the tunneling rate is finite, 
where the semiclassical action quadratically depends on the strength of the
magnetic field and linearly on the temperature.
Explicit expressions in various limits have been obtained, which
can be directly tested experimentally.
For example, 
for the system of electrons on a helium surface\cite{aomd}, the critical 
magnetic field for the dynamical localization is about 1 Tesla, and for the
heterojunction systems it is about 50 Tesla, depending on the material
parameters.

\noindent
{\bf Acknowledgments:}
This work was supported in part by Swedish NFR.

\end{document}